**Title:**
Direct visualization of the quantum vortex density law in rotating $^4$He

**Authors:**
C. Peretti[+], J. Vessaire[+], E. Durozoy[+], M. Gibert[+]

**Affiliation**:
[+] Univ. Grenoble Alpes, Institut Néel - CNRS UPR2940, 25 rue des Martyrs, BP 166, 38042 Grenoble Cedex 9, France.

**Summary**

The study of quantum vortex dynamics in He II holds great promise to refine quantum-fluid models. Bose-Einstein condensates, neutron stars or even superconductors exhibit quantum vortices, whose interactions are a key element of dissipation in these systems. These quantum objects have their velocity circulation around their core quantized and, in He II, a core as thin as a helium atom. They have been observed experimentally by indirect means, such as second sound attenuation or electron bubble imprints on photo sensitive material, and for the last twenty years, decorating cryogenic flows with particles has proved to be a powerful method to study these vortices. However, in these recent particle visualization observations, experimental stability, initial condition, stationarity and reproducibility are elusive or fragmented and 2d dynamical analysis are performed although most of the considered flows are inherently 3d. Here we show that we are able to visualize these vortices in the canonical and higher symmetry case of a stationary rotating superfluid bucket. Using direct visualization, we quantitatively verify Feynman's rule linking the resulting quantum vortex density to the imposed rotational speed. Our statistically meaningful results demonstrate that decorated quantum vortices behave as Feynman predicted. It follows that hydrogen flakes are good tracers of quantum vortices for stationary cases in He II. The observed vortex lattices are analogous to Abrikosov lattices found in superconductors and Bose-Einstein condensates. Moreover, these lattices aligned with the rotation axis can play the role of a well-defined and controlled initial condition for dynamical cases. We make the most of this stable configuration by observing collective wave mode propagation along quantum vortices and quantum vortex interactions in rotating He II. These results provide a new experimental baseline for models to evolve towards a better description of all quantum-fluids.

## Main text:
## Introduction

To model He II[1], one needs to use tools at the frontier between fluid and quantum mechanics. Up to this day, there is no unified theory able to describe the dynamics of He II like the Navier-Stokes equation does for classical viscous fluids. Nonetheless, we can assess that this phase of matter exhibits particular properties, such as the superposition of normal fluid and superfluid[2]. Additionally, the vorticity of the superfluid is constrained on quantum topological defects introduced by R. Feynman[3], called quantum vortices. They are atomically thin lines of macroscopic length, and are prime candidates to directly observe macroscopic objects displaying quantum properties. Quantum mechanics imposes that the circulation of the velocity field around these vortex lines is quantized: $\Gamma \equiv \oint v_s \cdot dl = n\kappa$, where $\kappa \equiv h/m \simeq 9.97 \times 10^{-8}$ m²/s is called the Feynman-Onsager quantum of circulation, $h$ is the Planck constant and $m$ the mass of a helium atom, $v_s$ is the superfluid velocity and $n$ is an integer. The velocity field around such a vortex is $|v_s| = \Gamma/2\pi r$ at distance $r$ from the vortex core. Considering kinetic energy, it is straightforward to demonstrate that the system will favor $n$ vortices with a circulation $\pm 1\kappa$ over a single vortex "carrying" $n$ quanta $\kappa$. Quantum vortices are an elementary brick of quantum-fluid systems, and the main subject of the present article. Experimentally, second sound tweezers[4–6], quartz tuning forks[7,8] and electrons (inside bubbles)[9] were extensively used to characterize quantum flows at relatively large scales (above the average intervortex spacing). More recently, micro-machined sensors such as cantilevers[10,11] and hotwires[12,13] allowed to lower the smallest scale accessible experimentally taking advantage of large-scale superfluid experimental facilities[14,15]. But so far none of these techniques identified any strong effect related to their interaction with a single vortex. To develop further models, Lagrangian probes (moving with the flow) entered the field inspired by classical fluid hydrodynamics[16,17]. Up to now, three families of particles have been used as Lagrangian probes: (a) solid particles at room temperature with a density that matches the fluid density[18] (hollow glass sphere) or non-buoyant nanoparticles[19], (b) hydrogen[20,21] or deuterium[22–24] flakes and (c) metastable $He_2^*$ molecule[25]. It is hard to conclude firmly on what the first two kinds of particles are tracking exactly[26–28], while the third kind traces the normal fluid component only above 1 K and could be trapped on vortices bellow 0.5 K[29]. Coming back to solid particles, the seminal work done in the early aughts[20] convinced the community that these particles could, under certain conditions, decorate quantum vortices[30–33]. To confirm this thoroughly, we needed a canonical and reproducible experiment that compares to existent predictions without any adjustable parameters. Inspired by the conceptual ease of experiments conducted in the seventies[34], we have designed an experiment able to bring this proof: a He II spinning bucket seeded with solid dihydrogen particles under controlled conditions[35]. In this article, we directly visualize the quantum vortex density law with di-hydrogen flakes, and we compare it quantitatively and without any adjustable parameter to Feynman's prediction. Their match constitutes our main evidence that the dihydrogen flakes are good tracers of quantum vortices in stationary regimes. Furthermore, we use this spinning superfluid bucket configuration as a well-defined and controlled initial condition to explore wave propagation along quantum vortices and vortex/vortex interaction.

### *The spinning superfluid bucket*

In a spinning superfluid bucket, as demonstrated experimentally in 1979[34], we expect that quantum vortices will form and align with the axis of rotation. Seen from the top, they arrange as a regular array that minimizes the free energy of the system while respecting the quantization of circulation of the superfluid[36]. This vortex array is characterized by the length scale δ, called the intervortex spacing. To estimate this length scale, it is worth noting that the vorticity of a normal fluid under solid body rotation is $2\vec{\Omega}$ everywhere. Therefore, if the number of quantum vortices per unit area in a plane perpendicular to $\vec{\Omega}$ is $n^* = 2\Omega/\kappa$, the circulation around any contour embracing many vortex lines is the same as the one obtained with the solid body rotation. The Feynman's rule simply follows, stating that $\delta \equiv 1/\sqrt{n^*} = \sqrt{\kappa/2\Omega}$. The hexagonal lattice characterized by this length scale $\delta$ is pictured on Figure 1c, and the primary goal of our experiment is to verify the Feynman's rule (or the quantum vortex density law in rotating $^4$He) that relates directly the rotation rate of the bucket to $\delta$ only through fundamental constants ($\kappa \equiv h/m$).

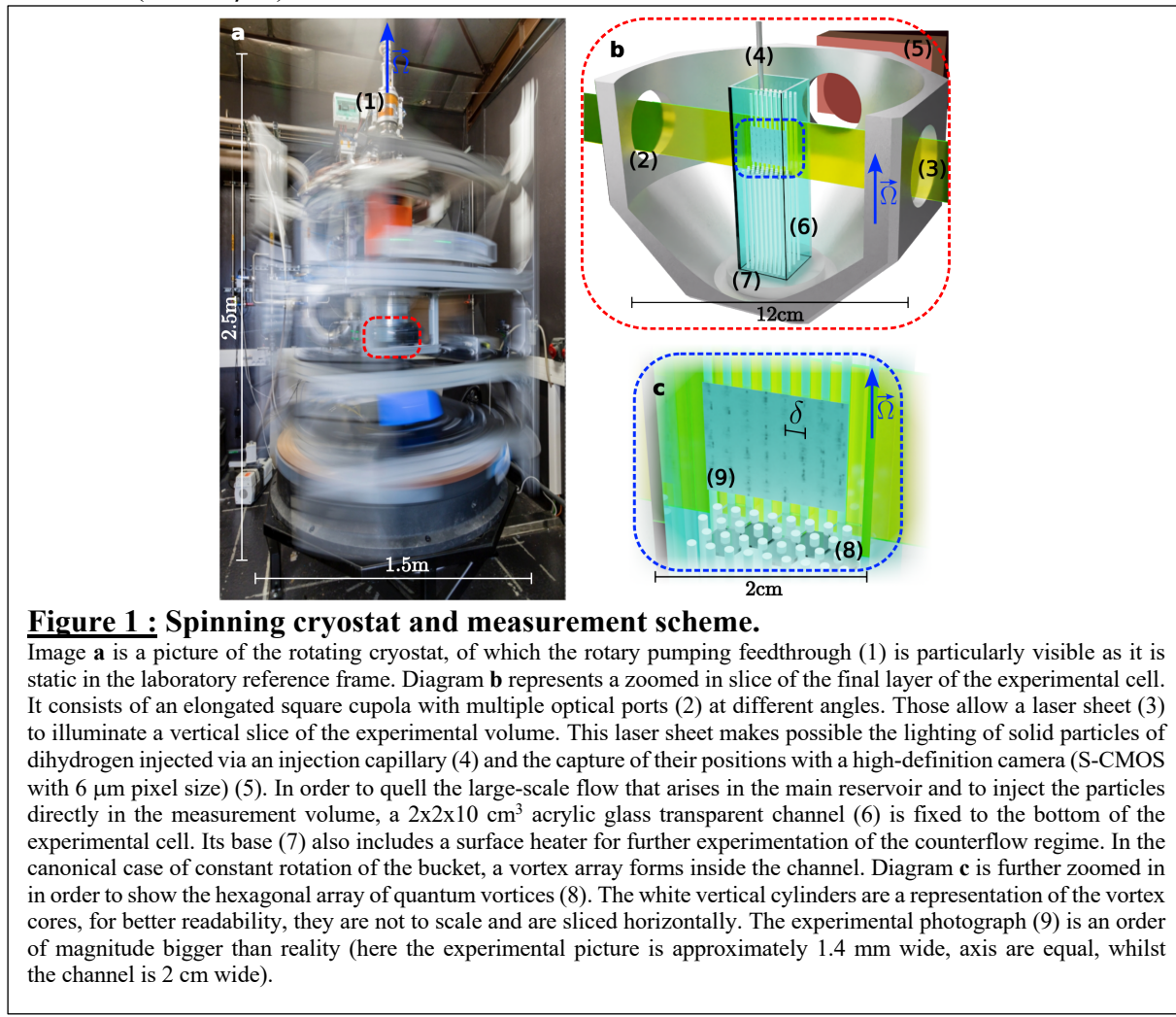

**Figure 1 : Spinning cryostat and measurement scheme.**
Image **a** is a picture of the rotating cryostat, of which the rotary pumping feedthrough (1) is particularly visible as it is static in the laboratory reference frame. Diagram **b** represents a zoomed in slice of the final layer of the experimental cell. It consists of an elongated square cupola with multiple optical ports (2) at different angles. Those allow a laser sheet (3) to illuminate a vertical slice of the experimental volume. This laser sheet makes possible the lighting of solid particles of dihydrogen injected via an injection capillary (4) and the capture of their positions with a high-definition camera (S-CMOS with 6 μm pixel size) (5). In order to quell the large-scale flow that arises in the main reservoir and to inject the particles directly in the measurement volume, a 2x2x10 cm³ acrylic glass transparent channel (6) is fixed to the bottom of the experimental cell. Its base (7) also includes a surface heater for further experimentation of the counterflow regime. In the canonical case of constant rotation of the bucket, a vortex array forms inside the channel. Diagram **c** is further zoomed in in order to show the hexagonal array of quantum vortices (8). The white vertical cylinders are a representation of the vortex cores, for better readability, they are not to scale and are sliced horizontally. The experimental photograph (9) is an order of magnitude bigger than reality (here the experimental picture is approximately 1.4 mm wide, axis are equal, whilst the channel is 2 cm wide).

*The spinning cryostat*

The spinning cryostat that we designed (see Figure 1a) in order to observe this vortex lattice is called CryoLEM for Cryogenic Lagrangian Exploration Module[37]. The experimental volume has the shape of an elongated square cupola that allows for multiple optical axis orientations when aiming at its center. The heat losses on this entire volume are below 100 mW, thanks to the thermal shield cooled with liquid nitrogen equipped with KG3 windows[38]. The spinning table main components are two 1.2 m diameter rectified cast-iron plates of about one ton each, see Figure 1a. The first one is stable in the laboratory and the second one levitates on a 50 μm air cushion carrying the spinning cryostat, plus equipment (laser, cameras, sensors and electronics). We use a pressurized air bearing to reduce mechanical vibrations to their lowest limits. On the axis of the table, in the bottom part, one can find the drive belt and the electrical feedthrough and, on the top, we have designed in house a rotary pumping feedthrough in order to perpetually control the pressure in the experimental liquid helium reservoir (hence the fluid temperature since we work on a saturated vapor bath). To visualize particles in our experimental volume, we use a 500 mW laser that we extend in one direction using cylindrical lenses arranged in an afocal telescope way. The light sheet enters the experiment chamber and leaves it through the windows in such a way that the laser beam generates almost no heating on the fluid (not noticeable on the pressure regulation system). In the middle of the experimental chamber, we have reduced even further the experimental volume by setting up a channel of 2x2 cm$^2$ square cross section and 10 cm high made out of acrylic transparent glass, as depicted in Figure 1b. The role of this channel is primarily to reduce the large-scale flow that arises in the main helium reservoir and to inject the particles directly in the measurement volume. The particles are created thanks to a 98 % helium, 2 % dihydrogen gas mixture at room temperature injected at the top of the channel through a 1.5mm diameter capillary. We have adopted an injection scheme[20] detailed in supplementary information Figure SI1.

*Measurement of the quantum vortex lattice*

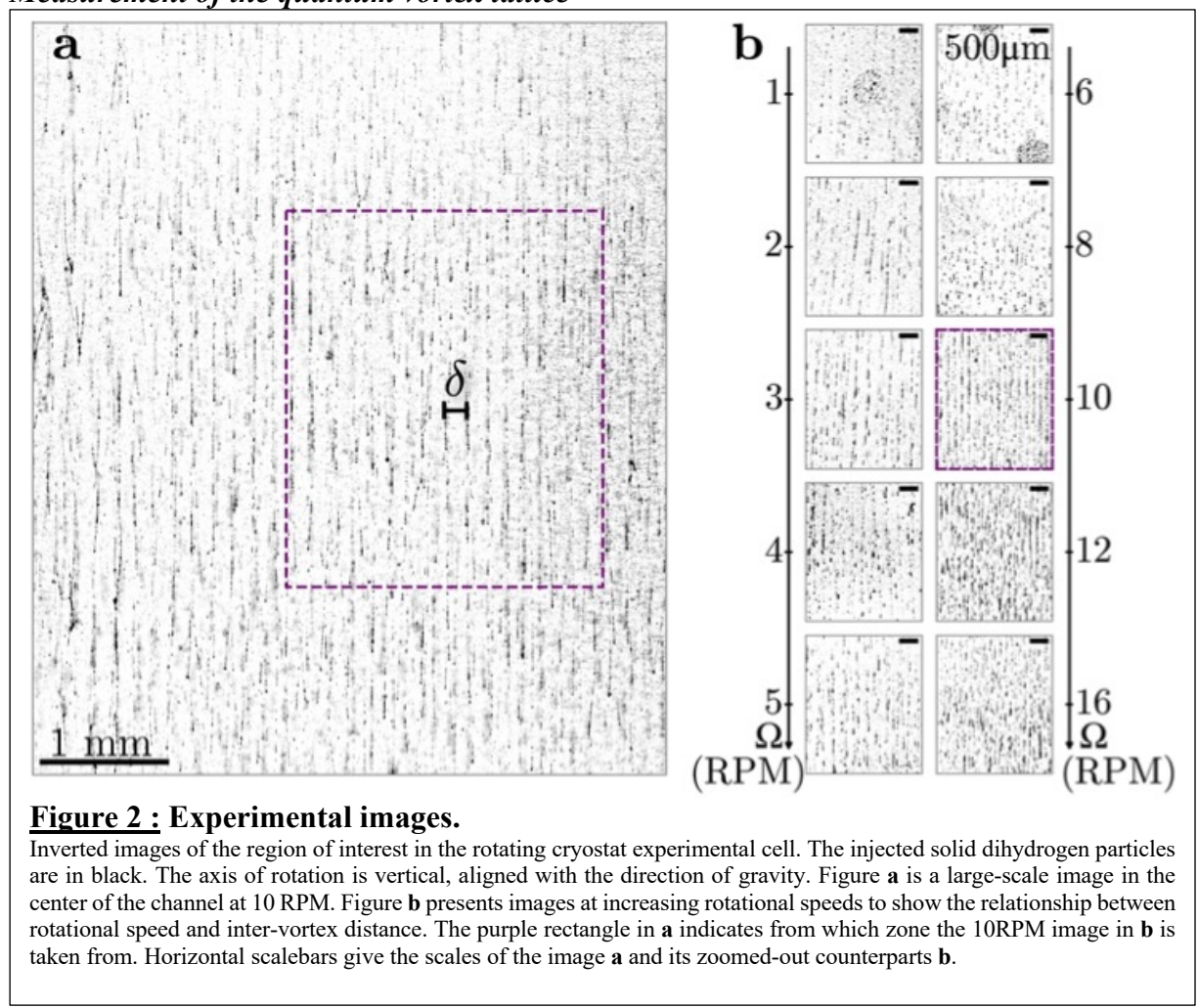

**Figure 2 : Experimental images.**
Inverted images of the region of interest in the rotating cryostat experimental cell. The injected solid dihydrogen particles are in black. The axis of rotation is vertical, aligned with the direction of gravity. Figure **a** is a large-scale image in the center of the channel at 10 RPM. Figure **b** presents images at increasing rotational speeds to show the relationship between rotational speed and inter-vortex distance. The purple rectangle in **a** indicates from which zone the 10RPM image in **b** is taken from. Horizontal scalebars give the scales of the image **a** and its zoomed-out counterparts **b**.

Before starting a new acquisition, the CryoLEM is allowed to spin at the considered rotation rate for about one hour and cooled down to about 2.2 K (just above the transition temperature). We inject the dihydrogen snowflakes and immediately after, start to cool down liquid helium down to 2.160 ± 0.005 K (see Figure SI1). We trigger the image acquisition when the liquid enters its He II phase. A few seconds after the transition, the turbulence triggered inside the region of interest has decayed and the diffuse cloud of particles seems to "condense" on the vortex lattice, as seen in Figure 2a (borders close to the boundaries were cropped) in which the injected solid di-hydrogen particles are in black. Since we visualize the lattice in a vertical plane that includes the axis of rotation, hydrogen particles appear as series of vertical pearl necklaces. This image offers a direct measurement of the intervortex spacing δ as the distance between two consecutive lines. The distance between vertical lines is reduced when rotational velocity increases as seen in Figure 2b. This visually demonstrates the expected consequence of Feynman's rule. The movies that we acquired from these patterns demonstrate a long-duration stability (see SI Movie 1). We have measured that the particles have an average upward velocity of $v_{settling} = 20 \pm 20$ μm/s that would correspond to particles with a diameter lower than 1 μm if one assumes the viscous Stokes drag with the normal fluid as unique restoring force. As we acquired more than 500 GB of images (~50000 frames), to analyze systematically this dataset, we have developed a technique using a complex Gabor wavelet detailed in the supplementary information. Using this method, we have measured for each experimental condition a series of separation reported on Figure 3 as a function of the rotation rate. The horizontal error bar corresponds to 2% uncertainty of the absolute value of the rotation rate and the vertical one is the dispersion measured by our algorithm (see supplementary information). On this figure we have also represented the Feynman's rule that has no adjustable parameter and describes properly our data.

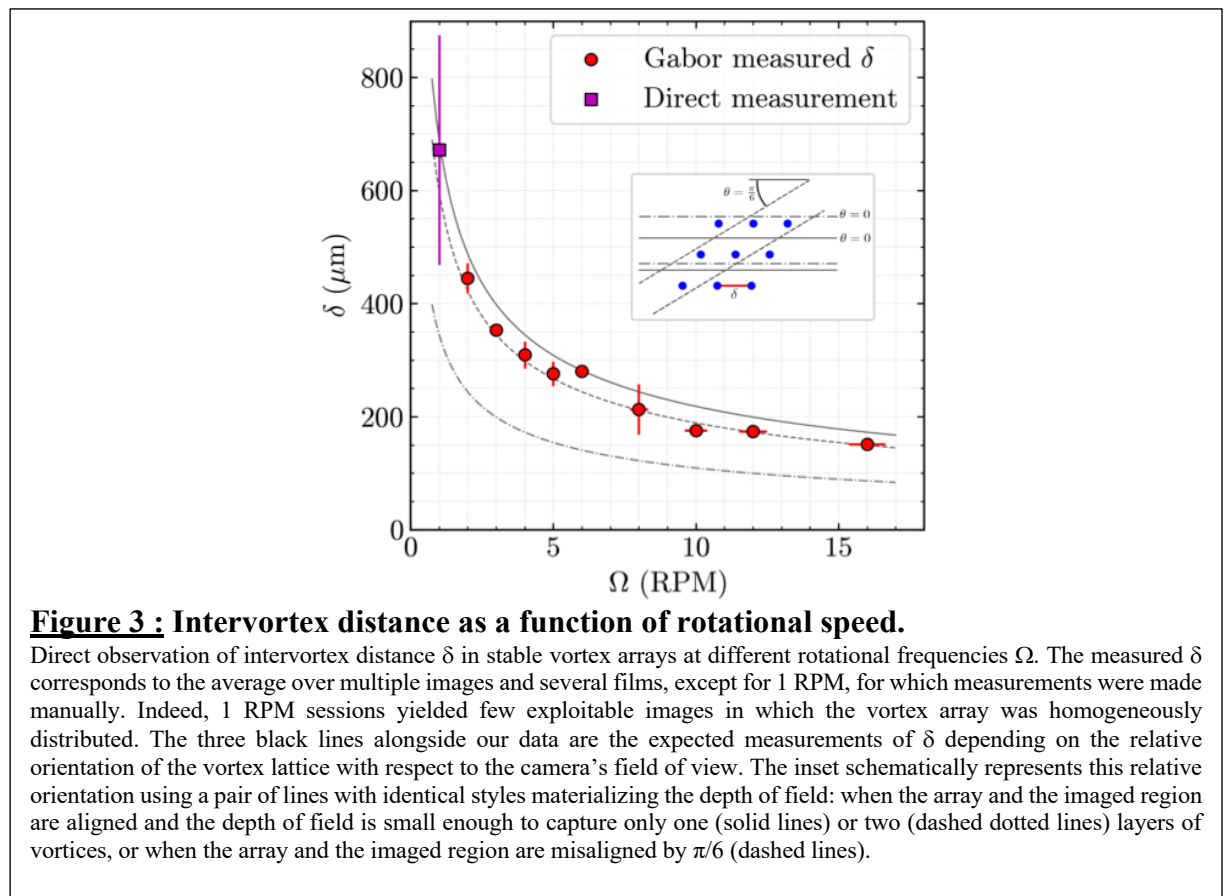

**Figure 3 : Intervortex distance as a function of rotational speed.**
Direct observation of intervortex distance δ in stable vortex arrays at different rotational frequencies Ω. The measured δ corresponds to the average over multiple images and several films, except for 1 RPM, for which measurements were made manually. Indeed, 1 RPM sessions yielded few exploitable images in which the vortex array was homogeneously distributed. The three black lines alongside our data are the expected measurements of δ depending on the relative orientation of the vortex lattice with respect to the camera's field of view. The inset schematically represents this relative orientation using a pair of lines with identical styles materializing the depth of field: when the array and the imaged region are aligned and the depth of field is small enough to capture only one (solid lines) or two (dashed dotted lines) layers of vortices, or when the array and the imaged region are misaligned by π/6 (dashed lines).

**Discussion**
The agreement between our measurements and the Feynman's rule over such a large data set analyzed using an objective mathematical approach is the main result of this paper. It demonstrates that direct visualization of quantum vortices via hydrogen flakes can lead to quantitative physical interpretations. Nevertheless, we have to discuss the fact that we have no way to align the imaged region with any of the vectors defining the unit cell of the expected hexagonal vortex lattice. The laser sheet enters the channel orthogonally to its wall and the quantum vortex lattice orients itself inside the channel. The boundary conditions of this canonical problem are such that about one layer of vortex lines is missing close to the border of the container because of the "image" vortex lines contribution to the velocity field. The effect of the square cross section of the channel on our measurement is out of the scope of this paper, but requires further research. Therefore, we are left with the fact that with a depth of field of about 150 μm, we can measure a separation between the exact Feynman's rule value $\delta$ if the principal axis of the lattice is colinear with the imaged region, and $\delta\,cos(\pi/6)$ if it is misaligned (see Figure 3 inset). All our data points are indeed in between those two limits. This allows to conclude that we have verified Feynman's rule by direct visualization of hydrogen solid particles trapped on the quantum vortices core. If we recall now that hydrogen flakes have a diameter of order 1 μm and that the quantum vortex core has a diameter of order 1 Å we could compare this situation to a *whale pod attached to a quantum fishing line*. This scale comparison recalls us that these particles may be passive in this particular stationary case as demonstrated by our measurements, but it may not be the case when considering a dynamical case. Furthermore, this lattice configuration is the perfect initial condition to trigger and study wave propagating along these quantum vortices and vortex/vortex interaction. At the bottom of the channel, we have installed a surface heater that can trigger a normal fluid momentum flux toward the open end of the channel. Below an amplitude of a few mW (out of the scope of this paper), nothing happens. Then, when this heater is driven with an alternating current, a collective wave mode is triggered along the vortex lines as shown on Figure 4. Its frequency is the one of the thermal excitations and its transverse amplitude (in horizontal displacement) depends on this frequency and exhibits a resonance when the normal fluid is driven vertically with a frequency equal to the rotation rate $\Omega$. These promising experimental results are here to "feed" different theoretical approaches. In particular, the initial threshold has to be linked to the Donnelly-Glaberson[39] instability, and the waves triggered on the lattice can be inertial[40], Kelvin or Tkachenko waves depending on the regime considered[41]. A thorough comparison between these experimental results and theoretical approaches[42] could lead to a proper description of the particle/vortex interaction and a better understanding of the value of the coefficients in Schwarz's equation[43].

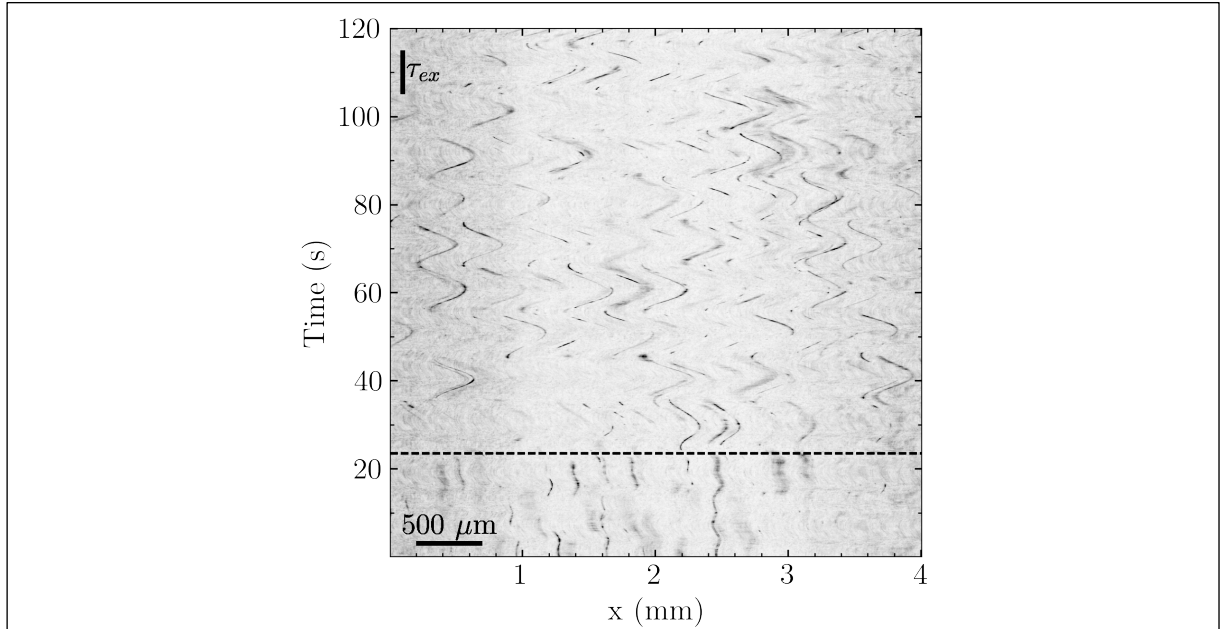

**Figure 4 : Space-time visualization of wave propagation along quantum vortices.**
Space-time representation of a single horizontal line of pixels in an experimental movie at rotational speed Ω=5 RPM. The horizontal scalebar represents a 500 μm scale, while the vertical scalebar $\tau_{ex}$ represents 10 s. The horizontal dotted line indicates the time $t_0$ after which the heater is activated with a periodic positive square function of period $\tau_{ex}$, and of heating power 3.6 mW over a period (9 W.m$^{-2}$). Although there exists periodic horizontal movement before $t_0$, the heat injection after $t_0$ generates ample periodic horizontal motion coming from the wave propagation along the vortices.

When the heater power is pushed even further, case considered in SI Movie 2, the quantum vortex lines start to "see" each other up to a point where they interact strongly and trigger thousands of interactions in between neighbouring vortices of the initial vortex lattice and in fine destroy it. The flow evolves toward a spinning counter flow state[44]. This opens a way to study the reconnection process between quantum vortices, systematically, from a well-known and established initial condition, while controlling the driving force. These reconnexions are expected to play a key role in the energy dissipation processes at ultra-low temperature but also in the establishment of the quantum vortex lattice[45]. To analyze these dynamical results further, a direct comparison with numerical simulations based on a model that fully couples the vortex line dynamics with the normal fluid, such as the recently developed FOUCAULT[46] model, is needed to account for finite temperature effects.

**Conclusion:**
In summary, through the use of a Gabor wavelet, we show that the spatial distribution of di-hydrogen flakes seeding a spinning He II bucket follows the Feynman's rule. Therefore, we experimentally demonstrated that di-hydrogen flakes are good tracers of quantum vortices in He II in stationary cases. This enables us to build reproducible and controlled initial conditions for further dynamical studies of quantum vortices. Additionally, we explore the effect of a surface heater driven by an alternating current at the bottom of our experimental channel. We've identified that below a certain threshold the vortex lattice remains unperturbed, and above, a collective wave mode arises. Preliminary results tend to prove that the amplitude of these waves is maximal when heat flux frequency equals the rotation rate of the experiment. When pushing the heat flux further, neighbor vortices start to see each other and interact. This opens a way to study quantum vortex reconnexion while controlling their initial condition and the driving force that triggers these extreme events. Both dynamical cases introduced here deserves further research that should lead to a better understanding of the physics of quantum vortices and ultimately improve superfluid models, thanks to direct visualization of this macroscopic quantum object in He II.

**Main references:**

**Tables:**

**Figure legends:**

**Methods:**

**Methods references:**

**Acknowledgements:**
We thank G. Garde for his deep implication in the project as lead engineer, his contribution to this work is crucial. We acknowledge P. Spathis , B. Chabaud and O. Bourgeois for our discussions all along the project. This work received the support of grants ANR-11-PDOC-0001 (3D-QuantumV), ANR-10-LABX-0051 (LANEF) and ANR-17-CE30-0003 (DisET).

**Author contributions:**
All authors significantly contributed to this work as a team effort. M.G. carries out the project since the beginning (2011). E.D. and J.V. defined and refined the protocols to control thoroughly the experiment, supervised by M.G.. J.V. and C.P. performed the experiments and their analysis, supervised by M.G.. M.G. wrote the first draft of the manuscript. All authors read critically and participate to the writing of the manuscript.

**Competing interest declaration:**
The authors declare no competing interests.

**Additional information:**
Supplementary information is available in the online version of the paper. Reprints and permissions information is available online at www.nature.com/reprints. Correspondence and requests for materials should be addressed to M.G.. The data that support the findings of this study are available from the corresponding author M.G. upon reasonable request.

**Extended data figure/table legends:**